\documentclass[12pt]{article}
\usepackage{latexsym,amssymb}
\usepackage[T1]{fontenc}
\usepackage[latin1]{inputenc}
\newcommand{\eps}      {\epsilon}
\newcommand{\R}        {\mathbb R}

\newcommand{\T}        {\mathbb T}
\newcommand{\A}        {\mathbb A}
\renewcommand{\AA}       {{\cal A}}
\renewcommand{\div}       {\hbox{{\rm div}}}
\newcommand{\curl}       {\hbox{{\rm curl}}}
\newcommand{\xeps}{\frac{x}{\eps}}
\newcommand{\dsp}{\displaystyle}
\newtheorem{theorem}      {Theorem}[section]
\newtheorem{theorem*}     {theorem}
\newtheorem{proposition}  [theorem]{Proposition}

\newtheorem{lemma}        [theorem]{Lemma}

\newtheorem{remark}       [theorem]{Remark}

\def\2s{\stackrel{\rm 2s}{\rightharpoonup}}
\title{Homogenization of the Schr\"{o}dinger
equation with a time oscillating potential}
\author{
Gr\'egoire Allaire\thanks{Centre de Math\'ematiques
Appliqu\'ees, Ecole Polytechnique, 91128 Palaiseau, France 
--- {\tt gregoire.allaire@polytechnique.fr}}
\and
M. Vanninathan\thanks{
TIFR Center, P.O. Box 1234, Bangalore - 560012, India ---
{\tt vanni@math.tifrbng.res.in}}
}

\begin{document}

\maketitle

\begin{abstract}
We study the homogenization of a Schr\"{o}dinger equation
in a periodic medium with a time dependent potential.
This is a model for semiconductors excited by an external
electromagnetic wave. We prove that, for a suitable choice
of oscillating (both in time and space) potential, one can
partially transfer electrons from one Bloch band to another.
This justifies the famous "Fermi golden rule" for the transition
probability between two such states which is at the basis
of various optical properties of semiconductors.
Our method is based on a combination of classical homogenization
techniques (two-scale convergence and suitable oscillating test
functions) and of Bloch waves theory.
\end{abstract}

\medskip

\begin{center}
Dedicated to the memory of Fr\'ed\'eric Poupaud.
\end{center}

\section{Introduction}

This work is devoted to the mathematical justification of a
problem of mean field approximation in solid state physics.
More precisely, we study the homogenization of the following
Schr\"{o}dinger equation
\begin{equation}
\label{eq1.1}
\left\{ \begin{array}{ll}
\dsp i \frac{\partial u_\eps}{\partial t} - \Delta u_\eps +
\left( \eps^{-2} c\left(\xeps\right) + d_\eps(t,x) \right) u_\eps = 0
& \mbox{ in } \R^N\times (0,T) \\[0.3cm]
u_\eps(t=0,x) = u_\eps^0(x) & \mbox{ in } \R^N ,
\end{array} \right.
\end{equation}
where $0<T<+\infty$ is a final time. The potential $c(y)$
is a real, bounded and periodic function defined for $y\in\T^N$
(the unit torus). Equation (\ref{eq1.1}) is the so-called
one-electron model for describing the electrons in a crystal 
or in a semiconductor, characterized by the periodic potential
$c(y)$ \cite{cohen}, \cite{cohen.che}, \cite{yu-cardona}.
An exterior field is also applied to the sample: in first
approximation it is described by another real potential
$d_\eps(t,x)$ which depends also on the time variable in
contrast to $c$ (see Section \ref{sec.em} for the case of an
electromagnetic potential). In (\ref{eq1.1}) the size of
the background microscopic potential is of order $\eps^{-2}$
while that of the exterior macroscopic potential is of order
$\eps^0$, so $d_\eps$ is just a small perturbation (the relative
size of which is the square of the period).
It can also be seen as a control acting on the semiconductor and
its specific dependence on $\eps$ will vary with the initial
condition and the desired final or target state at time $T$
(see (\ref{eq1.4}) below).
For example, the exterior potential can be light illuminating
the semiconductor: if its energy is high enough, it can excite
electrons from the valence band to conduction band. This effect
is called optical absorption. Its converse effect (light
emitted by electrons going from the conduction band to the
valence band) is at the root of many important devices
such as lasers, light emitting diodes and photo-detectors
\cite{cohen.che}, \cite{sapoval}, \cite{yu-cardona}.

Remark that Planck's constant has been normalized to unity in (\ref{eq1.1}).
We emphasize that the $\epsilon$-scaling in (\ref{eq1.1})
is not the usual semi-classical scaling for Schr\"{o}dinger
equation \cite{buslaev}, \cite{guillot}, \cite{gerard}, \cite{gerard2},
\cite{sjostrand}, \cite{pst} which would involve a $\epsilon^{-1}$
coefficient in front of the time derivative. Instead, it is the
scaling of homogenization as in \cite{acpsv}, \cite{ap}, \cite{blp}.
In physical terms it corresponds to an asymptotic study for
much longer times than in the semi-classical limit (we refer
to \cite{ap} for a more complete discussion of the scaling).

Let us describe a typical example of our results.
We introduce the so-called Bloch or shifted cell problem,
$$
- (\div_y +2i\pi\theta) \Big( (\nabla_y +2i\pi\theta)\psi_n \Big) + 
c(y) \psi_n = \lambda_n(\theta) \psi_n  \quad \mbox{ in } \T^N ,
$$
where $\theta\in\T^N$ is a parameter and $(\lambda_n(\theta),
\psi_n(y,\theta))$ is the $n$-th eigencouple. 
In physical terms, the Bloch frequency $\theta$ is
the quasi momentum and the range of $\lambda_n(\theta)$, 
as $\theta$ varies, is a Bloch or energy band.
We consider an initial data which is a wave packet of the type
\begin{equation}
\label{eq1.2}
u_\eps^0(x) = \psi_n\left(\xeps,\theta^n\right)
e^{2i\pi\frac{\theta^n\cdot x}{\eps}} v^0(x) ,
\end{equation}
and we would like to attain a final state at a different 
momentum $\theta^m$ and energy $\lambda_m(\theta^m)$
\begin{equation}
\label{eq1.3}
u_\eps^T(x) = \psi_m\left(\xeps,\theta^m\right) 
e^{2i\pi\frac{\theta^m\cdot x}{\eps}} v^T(x) .
\end{equation}
For achieving this goal we choose an oscillating potential
\begin{equation}
\label{eq1.4}
d_\eps(t,x) = \Re\left( e^{i\frac{(\lambda_m(\theta^m)-\lambda_n(\theta^n)) t}{\eps^2}}
e^{2i\pi\frac{(\theta^m-\theta^n)\cdot x}{\eps}} \right) d\left(t,x,\frac{x}{\eps}\right) ,
\end{equation}
where $\Re$ denotes the real part, and $d(t,x,y)$ is a real potential
defined on $[0,T]\times\R^N\times\T^N$. Formula (\ref{eq1.4}) for the 
potential is quite natural: the oscillating phase in time corresponds
to the energy difference between the initial and final state (\ref{eq1.2}), 
(\ref{eq1.3}), while the oscillating phase in space corresponds
to the momentum difference. In other words, the potential puts energy
and momentum in the system so as to have global conservation of these
quantities. 

In truth we cannot reach (even approximately) the desired final state
(\ref{eq1.3}). Instead we end up with a mixed state, combination of
(\ref{eq1.2}) and (\ref{eq1.3}). Under the assumption that
$\theta^n$ and $\theta^m$ are critical points of their
non-degenerate energy levels (which implies that the group velocities 
of the wave packets (\ref{eq1.2}) and (\ref{eq1.3}) vanish) and are
non-resonant (which means that no other state are excited by the
external potential), see (\ref{a1}) and (\ref{a2}) below, 
we shall prove in Theorem \ref{thm.main} that the
solution of (\ref{eq1.1}) satisfies
$$
u_\eps(t,x) \approx e^{i\frac{\lambda_n(\theta^n)t}{\eps^{2}}} 
e^{2i\pi\frac{\theta^n\cdot x}{\eps}}
\psi_n\left(\xeps,\theta^n\right) v_n(t,x) + 
e^{i\frac{\lambda_m(\theta^m)t}{\eps^{2}}}
e^{2i\pi\frac{\theta^m\cdot x}{\eps}} 
\psi_m\left(\xeps,\theta^m\right) v_m(t,x) ,
$$
where the macroscopic profile $(v_n,v_m)$ is the unique solution of the
following Schr\"{o}dinger homogenized coupled system
\begin{equation}
\label{eq1.7}
\left\{ \begin{array}{ll}
\dsp i \frac{\partial v_n}{\partial t} - \div \left( A^*_n \nabla v_n \right) 
+ d^*_{nm}(t,x) \, v_m = 0
& \mbox{ in } \R^N\times (0,T) \\[0.2cm]
\dsp i \frac{\partial v_m}{\partial t} - \div \left( A^*_m \nabla v_m \right)
+ d^*_{mn}(t,x) \, v_n = 0
& \mbox{ in } \R^N\times (0,T) \\[0.2cm]
v_n(t=0,x) = v^0(x) & \mbox{ in } \R^N \\
v_m(t=0,x) = 0 & \mbox{ in } \R^N ,
\end{array} \right. 
\end{equation}
with homogenized coefficients $A^*_n, A^*_m$ and $d^*_{nm}=\overline d^*_{mn}$.
The tensors $A^*_n$ and $A^*_m$ are the inverses of the
effective masses of the particles corresponding to the
initial and desired final state. The coupling coefficient
$d^*_{nm}$ is given by a formula (see (\ref{eq2s.7b}) and
Remark \ref{rem.fermi}) known as "Fermi golden rule"
\cite{cohen}, \cite{cohen-tan}, \cite{sapoval}. The homogenized
system (\ref{eq1.7}) is a model for light absorption in
semiconductors. When $\theta^n=\theta^m$ one talks about
"direct" absorption, and when $\theta^n\neq\theta^m$ about
"indirect" absorption \cite{cohen}, \cite{cohen.che},
\cite{sapoval}, \cite{yu-cardona}. In truth, one does
not find (\ref{eq1.7}), as it stands, in the physical 
literature where instead a simpler semi-classical picture
is used. Specifically, physicists talk about the transition
probability between the two states (\ref{eq1.2}) and
(\ref{eq1.3}), which is precisely equal to the squared
modulus of the coupling coefficient $d^*_{nm}$.

To obtain the homogenized limit (\ref{eq1.7}) we follow
the method introduced in \cite{acpsv}, \cite{ap}. 
The main idea is to
use Bloch wave theory to build adequate oscillating test
functions and pass to the limit using two-scale
convergence \cite{allaire}, \cite{nguetseng}. 

The content of this paper is as follows. Section \ref{sec.notation} 
is devoted to recalling basic facts about Bloch waves and
two-scale convergence as well as stating our main assumption. 
Section \ref{sec.main} gives our main result of homogenization.
Section \ref{sec.em} focuses on a generalization of the 
Schr\"{o}dinger equation (\ref{eq1.1}) which takes into account
an electromagnetic field. Finally Section \ref{sec.reso} is
concerned with a resonant case where more than two states are
coupled.

\section{Bloch spectrum and two-scale convergence}
\label{sec.notation}

In this section we recall some results on Bloch
waves and two-scale convergence, and we introduce
our main assumptions on the initial and target states.

We assume that the potential $c(y)$ is a real measurable
bounded periodic function, i.e. belongs to $L^\infty(\T^N)$,
while the potential $d(t,x,y)$, appearing in (\ref{eq1.4}), is real,
measurable, uniformly bounded, periodic in $y$ and smooth in $(t,x)$.
We recall that, for given $\theta$, the Bloch (or shifted) spectral cell equation
\begin{equation}
\label{eq3.2}
- (\div_y +2i\pi\theta) \Big( (\nabla_y +2i\pi\theta)\psi_n \Big) +
c(y) \psi_n = \lambda_n(\theta) \psi_n  \quad \mbox{ in } \T^N ,
\end{equation}
admits a countable sequence of real increasing eigenvalues
$(\lambda_n)_{n\geq1}$ (repeated with their multiplicity)
and normalized eigenfunctions $(\psi_n)_{n\geq1}$,
with $\|\psi_n\|_{L^2(\T^N)}=1$, since its Green operator
is a compact self-adjoint complex-valued operator on $L^2(\T^N)$.
The dual parameter $\theta$ is called the Bloch frequency or
quasi momentum and it runs in the dual cell of $\T^N$, which, 
by our choice of normalization factor $2\pi$ in the phase 
factor, is again the unit torus $\T^N$. In other words, 
by periodicity it is enough to consider $\theta\in\T^N$. 
For more details on Bloch waves, see e.g. \cite{cpvbook},
\cite{kuchment}, \cite{reedsimon}.

In the sequel, we shall consider two energy levels $n,m\geq1$ and
Bloch parameters $\theta^n,\theta^m\in\T^N$ such that the eigenvalues
$\lambda_n(\theta^n)$ and $\lambda_m(\theta^m)$ satisfy the
following assumption
\begin{equation}
\label{a1}
\mbox{for } p=n,m \
\left\{ \begin{array}{ll}
(i) & \lambda_p(\theta^p) \mbox{ is a simple eigenvalue,} \\
(ii) & \theta^p \mbox{ is a critical point of } \lambda_p(\theta)
\mbox{ i.e., } \nabla_\theta\lambda_p(\theta^p) = 0 .
\end{array} \right.
\end{equation}
Of course, we assume that, either $n\neq m$, or $n=m$ and
$\theta^n\neq\theta^m$ (if $n=m$ and $\theta^n=\theta^m$,
then $d_\eps=d(t,x,x/\eps)$ and this case was already
treated in \cite{ap}). The simplicity assumption, i.e. 
part $(i)$ of (\ref{a1}), is generic and simplifies considerably 
the analysis. In particular, it implies that $\lambda_p(\theta)$ 
is infinitely differentiable in a vicinity of $\theta^p$, and 
one can introduce the group velocity $\nabla_\theta\lambda_p(\theta)$. 
The criticality assumption, i.e. part $(ii)$ of (\ref{a1}), 
is physically relevant when states at the bottom or top of 
Bloch bands are considered. For a discussion of this type 
of assumptions, as well as possible weaker ones, we refer 
to \cite{ap}. 

We also add a non-resonant assumption
\begin{equation}
\label{a2}
\begin{array}{ll}
\begin{array}{ll}
(iii) & \mbox{ for any } p\geq1, \quad
\lambda_p(2\theta^n-\theta^m)\neq 2\lambda_n(\theta^n)- \lambda_m(\theta^m)  .
\end{array}
\end{array}
\end{equation}
The interpretation of assumption (\ref{a2}) is the following.
The oscillating potential $d_\eps$, defined by (\ref{eq1.4}),
has been designed to transfer the initial state with (quasi)
momentum $\theta^n$ and energy $\lambda_n(\theta^n)$ to the target
state $\theta^m,\lambda_m(\theta^m)$. The only requirement
is that momentum and energy are conserved during this process.
Actually there is another possible state that can be reached
under the conservative action of $d_\eps$, namely the state
with momentum $2\theta^n-\theta^m$ and energy
$2\lambda_n(\theta^n)- \lambda_m(\theta^m)$. In order to
simplify the analysis, assumption (\ref{a2}) forbids this
additional state as a standing wave solution of (\ref{eq1.1})
without exterior potential. Section \ref{sec.reso} explores 
the resonant case where (\ref{a2}) is not satisfied.

Under assumption (\ref{a1}) it is well-known \cite{kato} that
one can make a choice of an eigenvector such that the $n$-th
eigencouple of (\ref{eq3.2}) is smooth in a neighborhood of $\theta^n$.
Introducing the operator $\A_n(\theta)$ defined on $L^2(\T^N)$ by
\begin{equation}
\label{eq3.3}
\A_n(\theta)\psi =
- (\div_y +2i\pi\theta) \Big( (\nabla_y +2i\pi\theta)\psi \Big) +
c(y) \psi - \lambda_n(\theta) \psi ,
\end{equation}
we differentiate (\ref{eq3.2}) with respect to $\theta$.
Denoting by $(e_k)_{1\leq k\leq N}$ the canonical basis
of $\R^N$ and by $(\theta_k)_{1\leq k\leq N}$ the components
of $\theta$, the first derivative satisfies
\begin{equation}
\label{eq3.4}
\A_n(\theta)\frac{\partial\psi_n}{\partial\theta_k} = 
2i\pi e_k \cdot (\nabla_y +2i\pi\theta)\psi_n + 
(\div_y +2i\pi\theta) \left( 2i\pi e_k \psi_n \right) + 
\frac{\partial\lambda_n}{\partial\theta_k}(\theta) \psi_n ,
\end{equation}
and the second derivative is
\begin{equation}
\label{eq3.5}
\begin{array}{ll}
\dsp \A_n(\theta)\frac{\partial^2\psi_n}{\partial\theta_k\partial\theta_l} 
&\dsp = 2i\pi e_k \cdot (\nabla_y 
+2i\pi\theta)\frac{\partial\psi_n}{\partial\theta_l} + 
(\div_y +2i\pi\theta) \left( 2i\pi e_k 
\frac{\partial\psi_n}{\partial\theta_l} \right) \\[0.3cm]
&\dsp + 2i\pi e_l \cdot (\nabla_y 
+2i\pi\theta)\frac{\partial\psi_n}{\partial\theta_k} + 
(\div_y +2i\pi\theta) \left( 2i\pi e_l 
\frac{\partial\psi_n}{\partial\theta_k} \right) \\[0.3cm]
&\dsp + \frac{\partial\lambda_n}{\partial\theta_k}(\theta) 
\frac{\partial\psi_n}{\partial\theta_l} 
+ \frac{\partial\lambda_n}{\partial\theta_l}(\theta) 
\frac{\partial\psi_n}{\partial\theta_k}  \\[0.3cm]
&\dsp -8\pi^2e_k \cdot e_l \psi_n +
\frac{\partial^2\lambda_n}{\partial\theta_l\partial\theta_k}(\theta) \psi_n
\end{array}
\end{equation}
Under assumption (\ref{a1}) we have $\nabla_\theta \lambda_n(\theta^n)=0$, 
thus equations (\ref{eq3.4}) and (\ref{eq3.5}) simplify 
for $\theta=\theta^n$ and we find
\begin{equation}
\label{eq3.6}
\frac{\partial\psi_n}{\partial\theta_k}(\theta^n) = 2i\pi \zeta^k_n , \quad
\frac{\partial^2\psi_n}{\partial\theta_k\partial\theta_l}(\theta^n) =
-4\pi^2 \chi_n^{kl} ,
\end{equation}
where $\zeta^k_n$ satisfies
\begin{equation}
\label{eq4.4}
\A_n(\theta^n)\zeta^k_n = 
e_k \cdot (\nabla_y +2i\pi\theta^n)\psi_n + 
(\div_y +2i\pi\theta^n) \left( e_k \psi_n \right) 
\quad \mbox{ in } \T^N ,
\end{equation}
and $\chi_n^{kl}$ satisfies
\begin{equation}
\label{eq4.5}
\begin{array}{ll}
\dsp \A_n(\theta^n)\chi_n^{kl} =
&\dsp e_k \cdot (\nabla_y +2i\pi\theta^n)\zeta_n^l +
(\div_y +2i\pi\theta^n) \left( e_k \zeta_n^l \right) \\[0.3cm]
&\dsp + e_l \cdot (\nabla_y +2i\pi\theta^n)\zeta^k_n + 
(\div_y +2i\pi\theta^n) \left( e_l \zeta^k_n \right) \\[0.3cm]
&\dsp + 2 e_k \cdot e_l \psi_n - \frac{1}{4\pi^2}
\frac{\partial^2\lambda_n}{\partial\theta_l\partial\theta_k}(\theta^n) \psi_n 
\quad \mbox{ in } \T^N .
\end{array}
\end{equation}
We know that $\zeta^k_n$ and $\chi_n^{kl}$ exist since they are
defined by (\ref{eq3.6}) as multiple of the derivatives of $\psi_n$
with respect to $\theta$ (and assumption (\ref{a1}) ensures that
$\psi_n$ is indeed differentiable). However, if we forget for a
moment definition (\ref{eq3.6}), the existence and uniqueness
of the solutions to (\ref{eq4.4}) and (\ref{eq4.5}) is not at
all obvious. Since the operator $\A_n(\theta^n)$ has a non
empty kernel spanned by $\psi_n$, one should apply the Fredholm
alternative: equations (\ref{eq4.4}) and (\ref{eq4.5}) admit
a unique solution (up to the addition of a multiple of $\psi_n$)
if and only if their right hand side are orthogonal to $\psi_n$
(i.e. satisfy the Fredholm compatibility condition). This
compatibility condition is not immediately satisfied. Actually,
it gives new informations which are a consequence of the
previously established existence of $\zeta^k_n$ and $\chi_n^{kl}$.
In particular, the compatibility condition of (\ref{eq4.5})
yields a formula for the Hessian matrix
$\nabla_\theta\nabla_\theta \lambda_n(\theta^n)$
in terms of $\psi_n$ and $\zeta^k_n$ that we shall use
later (see (\ref{eq4.11b})).

\begin{remark}
\label{rem.genA}
All our results can be generalized if we replace the Laplacian
in (\ref{eq1.1}) by the more general operator
$\div (A(y) \nabla\cdot)$ where $A(y)$ is a symmetric, bounded,
periodic and uniformly coercive matrix. In this case, the Bloch
spectral cell problem (\ref{eq3.2}) becomes
$$
- (\div_y +2i\pi\theta) \Big( A(y) (\nabla_y +2i\pi\theta)\psi_n \Big) + 
c(y) \psi_n = \lambda_n(\theta) \psi_n  \quad \mbox{ in } \T^N .
$$
A tensor $A(y)\neq Id$ may be interpreted as a periodic metric. 
It makes sense for the study of wave propagation in a periodic
media (see e.g. \cite{acpsv}). 
\end{remark}

Finally we recall the notion of two-scale convergence
introduced in \cite{allaire}, \cite{nguetseng}.

\begin{proposition}
\label{prop2s}
Let $u_{\epsilon}$ be a sequence uniformly bounded in $L^2(\R^N)$.
There exists a subsequence, still denoted by $u_\epsilon$, and a
limit $u_0(x,y) \in L^2(\R^N\times\T^N)$ such that
$u_\epsilon$ {\em two-scale converges\/} (weakly) to $u_0$ in the
sense that
$$
\lim_{\epsilon\to 0}
\int_{\R^N} u_\epsilon(x)\phi(x,\xeps)\,dx =
\int_{\R^N}\int_{\T^N} u_0(x,y)\phi(x,y)\,dx\,dy
$$
for all functions $\phi(x,y)\in L^2\left( \R^N ; C_\#(\T^N) \right)$.
\end{proposition}

\noindent
{\bf Notation:} for any function $\phi(x,y)$ defined on $\R^N\times\T^N$, 
we denote by $\phi^\eps$ the function $\phi(x,\xeps)$.

\section{Main result}
\label{sec.main}

Due to our assumptions on the coefficients, if the initial data 
$u_\eps^0$ belongs to $H^1(\R^N)$, there exists a unique solution 
of the Schr\"{o}dinger equation (\ref{eq1.1}) in
$C\left([0,T];H^1(\R^N)\right)$ which satisfies
the following a priori estimate. 

\begin{lemma}
\label{lem.estim-s}
There exists a constant $C>0$, which depends on $T$ but not
on $\eps$, such that the solution of (\ref{eq1.1}) satisfies
\begin{equation}
\label{eq2s.5}
\begin{array}{lcl}
\dsp \| u_\eps \|_{L^\infty\left((0,T);L^2(\R^N)\right)} &=&
\dsp \| u^0_\eps \|_{L^2(\R^N)} , \\[0.1cm]
\dsp \eps \| \nabla u_\eps \|_{L^\infty\left((0,T);L^2(\R^N)^N 
\right)} 
&\leq & 
\dsp C \left( \| u^0_\eps \|_{L^2(\R^N)} + \eps
\| \nabla u^0_\eps \|_{L^2(\R^N)^N} \right) .
\end{array}
\end{equation}
\end{lemma}

\noindent
{\bf Proof of Lemma \ref{lem.estim-s}.}
We multiply equation (\ref{eq1.1}) by $\overline{u_\eps}$ and we 
integrate by parts. Since all coefficients are real, taking the 
imaginary part yields
$$
\frac{d}{dt} \int_{\R^N} |u_\eps(t,x)|^2 dx = 0 .
$$
Next we multiply (\ref{eq1.1}) by $\eps^2\frac{\partial\overline{
u_\eps}}{\partial t}$ 
and we take the real part to get
$$
\frac{d}{dt} \int_{\R^N} \left( \eps^2 |\nabla u_\eps|^2 + \left( c\left(\xeps\right) 
+ \eps^2 d_\eps(t,x) \right) |u_\eps|^2 \right) dx = 
- 2\eps^2 \int_{\R^N} \frac{\partial d_\eps}{\partial t} (t,x) 
|u_\eps|^2 \, dx ,
$$
which yields (\ref{eq2s.5}) since 
$\eps^2 \frac{\partial d_\eps}{\partial t}$ is bounded 
in view of (\ref{eq1.4}).
$\Box$

\bigskip

Our main result is the following homogenization theorem.

\begin{theorem}
\label{thm.main}
Assume (\ref{a1}) and (\ref{a2}) and that the initial data $u^0_\eps\in H^1(\R^N)$
is
\begin{equation}
\label{eq2s.6b}
u^0_\eps(x) = \psi_n\left(\frac{x}{\eps},\theta^n\right)
e^{2i\pi\frac{\theta^n\cdot x}{\eps}} v^0(x) ,
\end{equation}
with $v^0\in H^1(\R^N)$.
Then the solution of (\ref{eq1.1}) can be written as
\begin{equation}
\label{eq2s.6}
\begin{array}{l}
\dsp u_\eps(t,x) = e^{i\frac{\lambda_n(\theta^n)t}{\eps^{2}}}
e^{2i\pi\frac{\theta^n\cdot x}{\eps}} \psi_n\left(\frac{x}{\eps},\theta^n\right)
v_n(t,x) \\[0.2cm]
\dsp \phantom{u_\eps(t,x)} +
e^{i\frac{\lambda_m(\theta^m)t}{\eps^{2}}}
e^{2i\pi\frac{\theta^m\cdot x}{\eps}} \psi_m\left(\frac{x}{\eps},\theta^m\right)
v_m(t,x) + r_\eps(t,x) ,
\end{array}
\end{equation}
with
\begin{equation}
\label{eq2s.6c}
\lim_{\eps\to0} \int_0^T \int_{\R^N} \left| r_\eps(t,x) \right|^2 dx = 0 ,
\end{equation}
and $(v_n,v_m)\in C\left([0,T];L^2(\R^N)\right)^2$ is the unique solution
of the homogenized Schr\"{o}dinger system
\begin{equation}
\label{eq2s.7}
\left\{ \begin{array}{ll}
\dsp i \frac{\partial v_n}{\partial t} - \div \left( A^*_n \nabla v_n \right) 
+ d^*_{nm}(t,x) \, v_m = 0
& \mbox{ in } \R^N\times (0,T) \\[0.3cm]
\dsp i \frac{\partial v_m}{\partial t} - \div \left( A^*_m \nabla v_m \right)
+ d^*_{mn}(t,x) \, v_n = 0
& \mbox{ in } \R^N\times (0,T) \\[0.2cm]
v_n(t=0,x) = v^0(x) & \mbox{ in } \R^N \\[0.2cm]
v_m(t=0,x) = 0 & \mbox{ in } \R^N ,
\end{array} \right.
\end{equation}
with $A^*_p= \frac{1}{8\pi^2} \nabla_\theta\nabla_\theta \lambda_p(\theta^p)$,
for $p=n,m$, and
\begin{equation}
\label{eq2s.7b}
d^*_{nm}(t,x) = \overline d^*_{mn}(t,x) = \frac12
\int_{\T^N} d(t,x,y)\overline\psi_n(y,\theta^n)\psi_m(y,\theta^m) \, dy .
\end{equation}
\end{theorem}

\begin{remark}
\label{rem.fermi}
Formula (\ref{eq2s.7b}), giving the coupling coefficient $d^*_{nm}$,
is a version of the famous "Fermi golden rule" in quantum mechanics
or solid state physics \cite{cohen}, \cite{cohen-tan}, \cite{sapoval}.
More precisely, the squared modulus of $d^*_{nm}$ is called the
transition probability per unit time from state $n$ to $m$ and
its formula is Fermi golden rule (see e.g. Chapter 6 in \cite{sapoval}).
The inverse tensor $(A^*_n)^{-1}$ is called the effective mass of
the particle corresponding to the wave function $v_n$ \cite{cohen-tan},
\cite{kittel}, \cite{myers}. These effective coefficients not only
depend on the chosen periodic crystal (characterized by the
potential $c(y)$) but also on the energy level or Bloch band $n,m$
of the particle, and on the quasi momentum $\theta^{n,m}$. Effective
mass theorems were already obtained in \cite{ap}, chapter 4 of
\cite{blp}, \cite{pedersen}, \cite{pr}. However, the derivation
of the coupled system (\ref{eq2s.7}) and the justification of the
Fermi golden rule is new to the best of our knowledge.

Since $\theta^n$ and $\theta^m$ are not necessarily minimum points,
the tensors $A^*_n$ and $A^*_m$ can be neither definite nor positive.
Nevertheless, the homogenized problem (\ref{eq2s.7}) is still well
posed in $C([0,T];L^2(\R^N))^2$ (by using semi-group theory \cite{brezis}),
although its solution may not belong to $L^2((0,T);H^1(\R^N))^2$.

In general $d^*_{nm}$ does not vanish, so that there is indeed a
coupling between the two modes $n$ and $m$. Nevertheless, if
$d(t,x,y)\equiv d(t,x)$ in (\ref{eq1.4}) and $\theta^n=\theta^m$,
then, by orthogonality of the modes, we have $d^*_{nm}=0$.
\end{remark}

\begin{remark}
As already noticed in \cite{ap}, the scaling of (\ref{eq1.1})
is not the usual semi-classical scaling for Schr\"{o}dinger
equation \cite{buslaev}, \cite{guillot}, \cite{gerard}, \cite{gerard2},
\cite{sjostrand}. The actual scaling of (\ref{eq1.1})
means that we are interested in much longer times
than in the semi-classical limit.
\end{remark}

\begin{remark}
As already said in Remark \ref{rem.genA} all our results, 
including Theorem \ref{thm.main} can be generalized if we
replace the Laplacian by the more general operator 
$\div (A(y) \nabla\cdot)$ with a real symmetric, bounded, periodic
and uniformly coercive matrix $A(y)$. 
\end{remark}

\begin{remark}
Theorem \ref{thm.main} still holds true if the initial data 
is given by a combination of the two states 
$$
u^0_\eps(x) = \psi_n\left(\frac{x}{\eps},\theta^n\right)
e^{2i\pi\frac{\theta^n\cdot x}{\eps}} v^0_n(x) + 
\psi_m\left(\frac{x}{\eps},\theta^m\right)
e^{2i\pi\frac{\theta^m\cdot x}{\eps}} v^0_m(x)
$$
instead of (\ref{eq2s.6b}). Of course, it yields a 
non-zero initial data for $v_m$ in (\ref{eq2s.7}). 
\end{remark}

\noindent
{\bf Proof of Theorem \ref{thm.main}.}
This proof is in the spirit of our previous works \cite{acpsv}, \cite{ap}. 
Define two sequences
\begin{equation}
\label{eq4.32}
\begin{array}{l}
\dsp v^n_\eps(t,x) = u_\eps(t,x) e^{-i\frac{\lambda_n(\theta^n)t}{\eps^{2}}}
e^{-2i\pi\frac{\theta^n\cdot x}{\eps}} ,\\
\dsp v^m_\eps(t,x) = u_\eps(t,x) e^{-i\frac{\lambda_m(\theta^m)t}{\eps^{2}}} 
e^{-2i\pi\frac{\theta^m\cdot x}{\eps}} .
\end{array}
\end{equation}
Since $|v^n_\eps|=|v^m_\eps|=|u_\eps|$, by the a priori estimates of 
Lemma \ref{lem.estim-s} we have, for $p=n,m$, 
$$
\| v^p_\eps \|_{L^\infty\left((0,T);L^2(\R^N)\right)} + 
\eps \| \nabla v^p_\eps \|_{L^2((0,T)\times\R^N)}
\leq C ,
$$
and applying the compactness of two-scale convergence 
(see Proposition \ref{prop2s}), up to a subsequence,
for $p=n,m$, there exists a limit $w_p(t,x,y)
\in L^2\left((0,T)\times\R^N;H^1(\T^N)\right)$
such that $v^p_\eps$ and $\eps\nabla v^p_\eps$ two-scale 
converge to $w_p$ and $\nabla_y w_p$, respectively.
Similarly, by definition of the initial data, 
$v^n_\eps(0,x)$ two-scale converges to
$\psi_n\left(y,\theta^n\right) v^0(x)$ and 
$v^m_\eps(0,x)$ two-scale converges to 0 if 
$\theta^m\neq\theta^n$ and to $\psi_n\left(y,\theta^n\right) v^0(x)$ 
if $\theta^m=\theta^n$. 

\noindent
{\bf First step.}
We multiply (\ref{eq1.1}) by the complex conjugate of 
$$
\eps^2\phi(t,x,\xeps) e^{i\frac{\lambda_n(\theta^n)t}{\eps^{2}}}
e^{2i\pi\frac{\theta^n\cdot x}{\eps}} 
$$ 
where $\phi(t,x,y)$ is a smooth test function defined 
on $[0,T)\times\R^N\times\T^N$, with compact support 
in $(t,x)$ for fixed $y$. Integrating by parts this yields
$$
\begin{array}{ll}
\dsp -i\eps^2 \int_{\R^N} u^0_\eps \overline\phi^\eps
e^{-2i\pi\frac{\theta^n\cdot x}{\eps}} dx - i \eps^2 
\int_0^T \int_{\R^N} v_\eps^n \frac{\partial\overline\phi^\eps}
{\partial t} dt \, dx & \\[0.4cm]
\dsp + \int_0^T \int_{\R^N} (\eps\nabla + 2i\pi\theta^n) v_\eps^n 
\cdot (\eps\nabla - 2i\pi\theta^n) \overline\phi^\eps \,dt \,dx & \\[0.4cm]
\dsp + \int_0^T \int_{\R^N} (c^\eps -\lambda_n(\theta^n)+\eps^2d^\eps) 
v_\eps^n \overline\phi^\eps \,dt \,dx & =0 .
\end{array}
$$
Passing to the two-scale limit yields the variational formulation 
of 
$$
- (\div_y +2i\pi\theta^n) \Big( (\nabla_y +2i\pi\theta^n)w_n \Big) +
c(y)w_n = \lambda_n(\theta^n) w_n \quad \mbox{ in } \T^N .
$$
By the simplicity of $\lambda_n(\theta^n)$, this implies 
that there exists a scalar function $v_n(t,x)\in
L^2\left((0,T)\times\R^N\right)$ such that 
\begin{equation}
\label{eq4.33}
w_n(t,x,y) = v_n(t,x) \psi_n(y,\theta^n) .
\end{equation}
Replacing $n$ by $m$ in the previous argument, a similar result holds 
true for $w_m(t,x,y) = v_m(t,x) \psi_m(y,\theta^m)$.

\noindent
{\bf Second step.} 
We multiply (\ref{eq1.1}) by the complex conjugate of 
\begin{equation}
\label{eq4.39}
\Psi_\eps = e^{i\frac{\lambda_n(\theta^n)t}{\eps^{2}}} 
e^{2i\pi\frac{\theta^n\cdot x}{\eps}} \left( 
\psi_n(\xeps,\theta^n) \phi(t,x) + \eps \sum_{k=1}^N 
\frac{\partial\phi}{\partial x_k}(t,x) \zeta^k_n(\xeps) \right)
\end{equation}
where $\phi(t,x)$ is a smooth test function with compact support 
in $[0,T)\times\R^N$, and $\zeta^k_n(y)$ is the solution of 
(\ref{eq4.4}). Integrating by parts, the resulting computation 
was already done in \cite{ap} in the absence of the oscillating 
potential $d_\eps(t,x)$. We briefly recall it:  after some algebra,
and using the summation convention for the repeated index $k$, we obtain
\begin{equation}
\label{eq4.6}
\begin{array}{ll}
\dsp \int_{\R^N} \nabla u_\eps \cdot \nabla \overline \Psi_\eps dx = 
& \dsp \int_{\R^N} (\nabla + 2i\pi\frac{\theta^n}{\eps}) 
(\overline\phi v_\eps^n) \cdot (\nabla - 2i\pi\frac{\theta^n}{\eps}) 
\overline\psi_n^\eps \\[0.4cm]
& \dsp +\eps \int_{\R^N} (\nabla + 2i\pi\frac{\theta^n}{\eps}) 
(\frac{\partial\overline\phi}{\partial x_k} v_\eps^n) \cdot 
(\nabla - 2i\pi\frac{\theta^n}{\eps}) \overline\zeta^{k,\eps}_n \\[0.4cm]
& \dsp - \int_{\R^N} e_k \frac{\partial\overline\phi}{\partial x_k} 
v_\eps^n \cdot (\nabla - 2i\pi\frac{\theta^n}{\eps}) \overline\psi_n^\eps 
\\[0.4cm]
& \dsp + \int_{\R^N} (\nabla + 2i\pi\frac{\theta^n}{\eps}) 
(\frac{\partial\overline\phi}{\partial x_k} v_\eps^n) \cdot 
e_k \overline\psi_n^\eps \\[0.4cm]
& \dsp - \int_{\R^N} v_\eps^n \nabla 
\frac{\partial\overline\phi}{\partial x_k} \cdot e_k \overline\psi_n^\eps
\\[0.4cm] 
& \dsp - \int_{\R^N} v_\eps^n \nabla 
\frac{\partial\overline\phi}{\partial x_k} \cdot 
(\eps\nabla - 2i\pi\theta^n) \overline\zeta^{k,\eps}_n \\[0.4cm]
& \dsp + \int_{\R^N} \overline\zeta^{k,\eps}_n 
(\eps\nabla + 2i\pi\theta^n)v_\eps^n \cdot \nabla 
\frac{\partial\overline\phi}{\partial x_k}  
\end{array}
\end{equation}
A first simplification arises from the definition of $\psi_n$ 
which satisfies, for any smooth compactly supported test function $\Phi$, 
\begin{equation}
\label{eq4.7}
\int_{\R^N} (\nabla + 2i\pi\frac{\theta^n}{\eps}) \psi_n^\eps 
\cdot (\nabla - 2i\pi\frac{\theta^n}{\eps}) \overline\Phi 
+ \frac{1}{\eps^2}\int_{\R^N} (c^\eps-\lambda_n(\theta^n)) 
\psi_n^\eps \overline \Phi = 0 . 
\end{equation}
A second simplification comes from the definition of $\zeta^k_n$ 
\begin{equation}
\label{eq4.8}
\begin{array}{l}
\dsp \int_{\R^N} (\nabla + 2i\pi\frac{\theta^n}{\eps}) 
\zeta^{k,\eps}_n \cdot (\nabla - 2i\pi\frac{\theta^n}{\eps}) 
\overline\Phi 
+ \frac{1}{\eps^2}\int_{\R^N} (c^\eps-\lambda_n(\theta^n)) 
\zeta^{k,\eps}_n \overline \Phi = \quad \\[0.4cm]
\dsp \quad \eps^{-1} 
\int_{\R^N} (\nabla + 2i\pi\frac{\theta^n}{\eps}) 
\psi_n^\eps \cdot e_k \overline\Phi - \eps^{-1} 
\int_{\R^N} e_k \psi_n^\eps \cdot 
(\nabla - 2i\pi\frac{\theta^n}{\eps}) \overline\Phi .
\end{array}
\end{equation}
Combining (\ref{eq4.6}) with the other terms of the variational
formulation of (\ref{eq1.1}), we easily 
check that the first line of its right hand side cancels out 
because of (\ref{eq4.7}) with $\Phi=\overline\phi v^n_\eps$, 
and the next three lines cancel out because of (\ref{eq4.8}) 
with $\Phi=\frac{\partial\overline\phi}{\partial x_k} v^n_\eps$.
We keep the three last terms of (\ref{eq4.6}) which are bounded.
Finally, (\ref{eq1.1}) multiplied by $\overline\Psi_\eps$ yields
after simplification
\begin{equation}
\label{eq4.9}
\begin{array}{ll}
\dsp -i \int_{\R^N} u^0_\eps \overline\Psi_\eps(t=0) dx -
i \int_0^T \int_{\R^N} v^n_\eps 
\left( \overline\psi_n^\eps \frac{\partial\overline\phi}{\partial t} + 
\eps \frac{\partial^2\overline\phi}{\partial x_k\partial t}
\overline\zeta^{k,\eps}_n \right) dt \, dx  & \\[0.4cm]
\dsp - \int_0^T \int_{\R^N} v^n_\eps \nabla 
\frac{\partial\overline\phi}{\partial x_k} \cdot e_k \overline\psi_n^\eps
 dt \, dx & \\[0.4cm] 
\dsp - \int_0^T \int_{\R^N} v^n_\eps \nabla 
\frac{\partial\overline\phi}{\partial x_k} \cdot 
(\eps\nabla - 2i\pi\theta^n) \overline\zeta^{k,\eps}_n dt \, dx &\\[0.4cm]
\dsp + \int_0^T \int_{\R^N} \overline\zeta^{k,\eps}_n 
(\eps\nabla + 2i\pi\theta^n)v^n_\eps \cdot \nabla 
\frac{\partial\overline\phi}{\partial x_k}  dt \, dx &\\[0.4cm]
\dsp + \int_0^T \int_{\R^N} d^\eps v^n_\eps \overline\Psi_\eps \, dt \, dx 
& =0 .
\end{array}
\end{equation}
We can pass to the two-scale limit in each term of (\ref{eq4.9})
as was done in \cite{ap}, except for the last one which is the 
only new and different term. The last line of (\ref{eq4.9}) 
is equal to
$$
\begin{array}{l}
\dsp \int_0^T \int_{\R^N} d\left(t,x,\frac{x}{\eps}\right) \frac12
\left( e^{i\frac{(\lambda_m(\theta^m)-2\lambda_n(\theta^n)) t}{\eps^2}}
e^{2i\pi\frac{(\theta^m-2\theta^n)\cdot x}{\eps}} +
e^{-i\frac{\lambda_m(\theta^m) t}{\eps^2}}
e^{-2i\pi\frac{\theta^m\cdot x}{\eps}} \right) \\
\dsp \phantom{\int_0^T \int_{\R^N}}
u_\eps \left( \overline\psi_n(\xeps,\theta^n) \overline\phi(t,x) + {\cal O}(\eps)
\right) \, dt \, dx \\
\dsp = \int_0^T \int_{\R^N} d\left(t,x,\frac{x}{\eps}\right) \frac12 \left(
v^{2n-m}_\eps + v^m_\eps \right)
\left( \overline\psi_n(\xeps,\theta^n) \overline\phi(t,x) + {\cal O}(\eps)
\right) \, dt \, dx
\end{array}
$$
where we introduced a new sequence $v^{2n-m}_\eps$ defined,
similarly to (\ref{eq4.32}), by
\begin{equation}
\label{eq4.23}
v^{2n-m}_\eps(t,x) = u_\eps(t,x)
e^{-i\frac{(2\lambda_n(\theta^n)-\lambda_m(\theta^m)) t}{\eps^2}}
e^{-2i\pi\frac{(2\theta^n-\theta^m)\cdot x}{\eps}} .
\end{equation}
Applying the same arguments as in the first step, this sequence
$v^{2n-m}_\eps$ is easily shown to two-scale converge to
$w_{2n-m}(t,x,y)$ which satisfies
\begin{equation}
\label{eq4.9b}
\begin{array}{l}
\dsp - (\div_y +2i\pi(2\theta^n-\theta^m)) \Big( 
(\nabla_y +2i\pi(2\theta^n-\theta^m)) w_{2n-m} \Big) \\[0.2cm]
\dsp + c(y)w_{2n-m} = (2\lambda_n(\theta^n)-\lambda_m(\theta^m))
w_{2n-m} \quad \mbox{ in } \T^N .
\end{array}
\end{equation}
Because of the non-resonance assumption (\ref{a2}), 
namely that $2\lambda_n(\theta^n)-\lambda_m(\theta^m)$ is not
equal to any eigenvalue $\lambda_p(2\theta^n-\theta^m)$,
the spectral problem (\ref{eq4.9b}) has no solution other than 0,
which implies that $w_{2n-m}(t,x,y)\equiv0$.
The two-scale limit of (\ref{eq4.9}) is thus
\begin{equation}
\label{eq4.10}
\begin{array}{ll}
\dsp -i \int_{\R^N} \int_{\T^N} |\psi_n|^2 v^0
\overline\phi(t=0) \, dx\, dy
- i \int_0^T \int_{\R^N} \int_{\T^N} |\psi_n|^2 v_n
\frac{\partial\overline\phi}{\partial t} dt \, dx \, dy & \\[0.4cm]
\dsp - \int_0^T \int_{\R^N} \int_{\T^N} \psi_n v_n \nabla 
\frac{\partial\overline\phi}{\partial x_k} \cdot e_k \overline\psi_n
 dt \, dx \, dy & \\[0.4cm] 
\dsp - \int_0^T \int_{\R^N} \int_{\T^N} \psi_n v_n \nabla 
\frac{\partial\overline\phi}{\partial x_k} \cdot 
(\nabla_y - 2i\pi\theta^n) \overline\zeta^k_n dt \, dx \, dy &\\[0.4cm]
\dsp + \int_0^T \int_{\R^N} \int_{\T^N} \overline\zeta^k_n 
(\nabla_y + 2i\pi\theta^n)\psi_n v_n \cdot \nabla 
\frac{\partial\overline\phi}{\partial x_k} dt \, dx \, dy&\\[0.4cm]
\dsp + \frac12 \int_0^T \int_{\R^N} \int_{\T^N} d(t,x,y) \psi_m v_m \overline\psi_n
\overline\phi \, dt \, dx  \, dy& =0 .
\end{array}
\end{equation}
To simplify (\ref{eq4.10}) we recall that $\int_{\T^N} |\psi_n|^2 dy =1$,
that $d^*_{nm}(t,x)$ is defined by (\ref{eq2s.7b}), and we introduce
\begin{equation}
\label{eq4.11b}
\begin{array}{ll}
\dsp 2 \left(A^*_n\right)_{jk} = & \dsp \int_{\T^N} \Big(
\psi_n e_j \cdot e_k \overline\psi_n +
\psi_n e_k \cdot e_j \overline\psi_n \\[0.3cm]
& \dsp + \psi_n e_j \cdot (\nabla_y - 2i\pi\theta^n) \overline\zeta^k_n 
+ \psi_n e_k \cdot (\nabla_y - 2i\pi\theta^n) \overline\zeta_n^j \\[0.3cm]
& \dsp - \overline\zeta^k_n (\nabla_y + 2i\pi\theta^n)\psi_n \cdot e_j
- \overline\zeta_n^j (\nabla_y + 2i\pi\theta^n)\psi_n \cdot e_k
\Big) dy .
\end{array}
\end{equation}
Because of the Fredholm compatibility condition of equation
(\ref{eq4.5}) for the second derivative of $\psi_n$, the matrix
$A^*_n$, defined by (\ref{eq4.11b}), is actually equal to
$\frac{1}{8\pi^2} \nabla_\theta\nabla_\theta \lambda_n(\theta^n)$.
Finally (\ref{eq4.10}) is equivalent to
\begin{equation}
\label{eq4.11}
\begin{array}{l}
\dsp -i \int_{\R^N} v^0 \overline\phi(t=0) \, dx -
i \int_0^T \int_{\R^N} v_n \frac{\partial\overline\phi}{\partial t} dt \, dx
- \int_0^T \int_{\R^N} A^*_n v_n \cdot \nabla \nabla\overline\phi dt \, dx \\
\dsp \phantom{i \int_{\R^N} v^0 \overline\phi(t=0) dx }
+ \int_0^T \int_{\R^N} d^*_{nm}(t,x) v_m \overline\phi \, dt \, dx =0 .
\end{array}
\end{equation}
A symmetric argument works for $v_m$ (changing $n$ in $m$ in the 
test function $\Psi_\eps$). However, the initial condition is zero 
in the homogenized equation for $v_m$. Indeed, either $\theta^m\neq\theta^n$ 
and we already know that $v_\eps^m(0)$ two-scale converges to 0, 
or $\theta^m=\theta^n$ and the orthogonality property 
$$
\int_{\T^N} \psi_n \overline\psi_m \, dy = 0 \quad\mbox{for}\quad m\neq n
$$
implies that the coefficient in front of the test function $\phi(0)$ 
vanishes, which in the variational formulation implies that $v_m(0)=0$. 
Therefore, (\ref{eq4.11}) and its counterpart for $m$ instead of $n$ 
yield a very weak form of the homogenized
system (\ref{eq2s.7}). Since $A^*_n, A^*_m$ are real symmetric
matrices and $d^*_{nm}=\overline d^*_{mn}$, (\ref{eq2s.7})
admits a unique solution in $C\left([0,T];L^2(\R^N)\right)^2$.
By uniqueness of this solution, the entire sequence $v_\eps^p$
two-scale converges weakly to $\psi_p\left(y,\theta^p\right) v_p(t,x)$
for $p=n,m$.

It remains to prove the strong convergence (\ref{eq2s.6c}).
We compute
\begin{equation}
\label{eq4.12}
\begin{array}{l}
\dsp
\| r_\eps(t) \|^2_{L^2(\R^N)} = \| u_\eps(t) \|^2_{L^2(\R^N)}
+ \|\psi_n^\eps v_n(t) \|^2_{L^2(\R^N)} +  \|\psi_m^\eps v_m(t) \|^2_{L^2(\R^N)}\\[0.2cm]
\dsp \phantom{\| r_\eps(t) \|^2_{L^2(\R^N)}}
- 2 {\cal R} \int_{\R^N} v^n_\eps(t) \overline\psi_n^\eps \overline v_n(t) \, dx
- 2 {\cal R} \int_{\R^N} v^m_\eps(t) \overline\psi_m^\eps \overline v_m(t) \, dx\\[0.2cm]
\dsp \phantom{\| r_\eps(t) \|^2_{L^2(\R^N)}}
+ 2 {\cal R} \int_{\R^N} e^{i\frac{(\lambda_n(\theta^n)-\lambda_m(\theta^m)) t}{\eps^2}}
e^{2i\pi\frac{(\theta^n-\theta^m)\cdot x}{\eps}} \psi_n^\eps v_n(t)
\overline\psi_m^\eps \overline v_m(t) \, dx .
\end{array}
\end{equation}
By the orthogonality property of the Bloch waves, the last integral in
(\ref{eq4.12}) converges to 0. By applying two-scale convergence, we
can pass to the limit in the second line and in the last two terms of
the first line of (\ref{eq4.12}). For the remaining term we use
Lemma \ref{lem.estim-s} which implies
$$
\| u_\eps(t) \|^2_{L^2(\R^N)} =
\| u^0_\eps \|^2_{L^2(\R^N)} \to \| \psi_n v^0 \|^2_{L^2(\R^N\times\T^N)}
= \| v^0 \|^2_{L^2(\R^N)}
$$
by the normalization condition of $\psi_n$. Thus we deduce
$$
\lim_{\eps\to0} \| r_\eps(t) \|^2_{L^2(\R^N)} =\| v^0 \|^2_{L^2(\R^N)}
-\| v_n(t) \|^2_{L^2(\R^N)} - \| v_m(t) \|^2_{L^2(\R^N)}
$$
which is precisely 0 because of the conservation of
energy of the homogenized system (\ref{eq2s.7}), i.e.
$$
\| v_n(t) \|^2_{L^2(\R^N)} + \| v_m(t) \|^2_{L^2(\R^N)} = \| v^0 \|^2_{L^2(\R^N)} .
$$
Since $\lim_{\eps\to0} \| r_\eps(t) \|^2_{L^2(\R^N)} = 0$, the Lebesgue
dominated convergence theorem yields (\ref{eq2s.6c}).
$\Box$

\begin{remark}
Recall that the function $\zeta_k(y)$ is the solution of (\ref{eq4.4}),
unique up to the addition of a multiple of $\psi_n$. This multiple
may depend on $(t,x)$ and therefore the test function $\Psi_\eps$,
as well as the homogenized system could depend on the choice of
this additive term. Actually the homogenized system depends on
$\zeta_k$ only through the homogenized tensor $A^*_n$, defined by
(\ref{eq4.11b}). If we replace $\zeta_k(y)$
by $\zeta_k(y) + c_k(t,x)\psi_n(y)$, an easy calculation shows
that all terms $c_k$ cancel out because of the Fredholm alternative
for $\zeta_k$, i.e. the right-hand side of (\ref{eq4.4}) is
orthogonal to $\psi_n$. Thus, the homogenized system is uniquely
defined whatever the choice of the additive constant in $\zeta_k(y)$. 
\end{remark}

\begin{remark}
A formal two-scale asymptotic expansion (in the spirit of \cite{blp})
of the solution $u_\eps$ of (\ref{eq1.1}) would give
$$
\begin{array}{l}
\dsp u_\eps(t,x) \approx e^{i\frac{\lambda_n(\theta^n)t}{\eps^2}}
e^{2i\pi\frac{\theta^n\cdot x}{\eps}}
\left( \psi_n\left(\frac{x}{\eps},\theta^n\right) v_n(t,x)
+ \eps \sum_{k=1}^N
\frac{\partial v_n}{\partial x_k}(t,x) \zeta^k_n(\xeps) \right) \\
\dsp \phantom{u_\eps(t,x)} + e^{i\frac{\lambda_m(\theta^m)t}{\eps^2}}
e^{2i\pi\frac{\theta^m\cdot x}{\eps}}
\left( \psi_m\left(\frac{x}{\eps},\theta^m\right) v_m(t,x)
+ \eps \sum_{k=1}^N
\frac{\partial v_m}{\partial x_k}(t,x) \zeta^k_m(\xeps) \right) .
\end{array}
$$
As usual in periodic homogenization, this expansion suggests
the choice of the test function $\Psi_\eps$, in the proof of
Theorem \ref{thm.main}. Another possible interpretation of
$\Psi_\eps$ is as follows. The large $\eps^{-2}$ terms in
the variational formulation of (\ref{eq1.1}) cancel out because
of the equation satisfied by $\psi_n$. However, new terms of
order $\eps^{-1}$ appear because of the first order derivatives
of $\psi_n$. They are compensated in turn by the second order
derivatives of the corrector $\zeta^k_n$.
\end{remark}

\begin{remark}
\label{rem.multiple}
Part $(i)$ of assumption (\ref{a1}) states that the eigenvalues
$\lambda_n(\theta^n)$ and $\lambda_m(\theta^m)$ are simple.
This hypothesis is crucial in order to be able to differentiate
the spectral cell problem with respect to $\theta$. If one of
these eigenvalues is not simple then, as is well known, it is
not anymore differentiable, but merely directionally differentiable
(which is not enough for our purpose). So, we do not know how to
generalize Theorem \ref{thm.main} in the case of multiple eigenvalues.
There is one notable exception when one eigenvalue is of multiplicity,
say $p>1$, and there exists locally a labelling of the eigenvalues
and eigenvectors
in $p$ smooth branches. Note that it is a very strong assumption,
which is rarely meet in practice. Then, using an argument of
\cite{ap}, one can generalize Theorem \ref{thm.main} and obtain
a limit system similar to (\ref{eq2s.7}), with as many equations
as the repeated multiplicities of the eigenvalues $\lambda_n(\theta^n)$
and $\lambda_m(\theta^m)$, and coupled only by zero-order terms.
\end{remark}

\begin{remark}
\label{rem.drift}
Part $(ii)$ of assumption (\ref{a1}) states that the group velocities
vanish, $\nabla_\theta\lambda_n(\theta^n)=\nabla_\theta\lambda_m(\theta^m)=0$.
If it is not the case, then it induces a large drift of order $\epsilon^{-1}$
and the homogenized system (\ref{eq2s.7}) can be obtained only in a
moving frame of reference, following this large drift (see \cite{ap}
for more details). Therefore, if $\nabla_\theta\lambda_n(\theta^n)
\neq \nabla_\theta\lambda_m(\theta^m)$, one can not generalize
Theorem \ref{thm.main} since both initial and target states
move with large different speeds, so no coupling is possible
in the limit as $\epsilon$ goes to zero. In the case
$\nabla_\theta\lambda_n(\theta^n)=\nabla_\theta\lambda_m(\theta^m)\neq0$
it is technically possible to generalize Theorem \ref{thm.main}, 
following the argument of \cite{ap}, but this result would not make
much sense since it would assume that the exterior potential
$d_\epsilon(t,x)$ move with the same velocity, or at least is
macroscopically constant, which is usually not the case in 
physical applications.
\end{remark}

\section{Electromagnetic potential}
\label{sec.em}

Instead of (\ref{eq1.1}) we now consider a Schr\"{o}dinger equation 
with an exterior electromagnetic field
\begin{equation}
\label{eq5.1}
\left\{ \begin{array}{ll}
\dsp i \frac{\partial u_\eps}{\partial t} - \left(\div+i\eps \AA_\eps\right)
\left( \nabla+i\eps \AA_\eps\right) u_\eps + \eps^{-2} c\left(\xeps\right)
u_\eps = 0 & \mbox{ in } \R^N\times (0,T) \\[0.3cm]
u_\eps(t=0,x) = u_\eps^0(x) & \mbox{ in } \R^N ,
\end{array} \right. 
\end{equation}
where $\AA_\eps(t,x)$ is the electromagnetic vector potential, i.e.
a function from $\R^+\times\R^N$ into $\R^N$ \cite{cohen},
\cite{cohen-tan}, \cite{sapoval}. The electric field $E$ and
magnetic field $B$ are recovered by
$$
E(t,x) = - \frac{\partial \AA_\eps}{\partial t}(t,x) \quad\mbox{and}\quad
B(t,x) = \curl \AA_\eps(t,x) .
$$
For an electromagnetic wave, the vector potential is assumed to be given by
\begin{equation}
\label{eq5.4}
\AA_\eps(t,x) = {\cal R}\left( e^{i\frac{(\lambda_m(\theta^m)-\lambda_n(\theta^n)) t}{\eps^2}}
e^{2i\pi\frac{(\theta^m-\theta^n)\cdot x}{\eps}} \right) a\left(t,x,\frac{x}{\eps}\right) ,
\end{equation}
where ${\cal R}$ denotes the real part and $a(t,x,y)$ is a bounded
smooth function from $\R^+\times\R^N\times\T^N$ into $\R^N$.
As before, $c(y)$ is a bounded function from $\T^N$ into $\R$,
the initial data $u_\eps^0$ belongs to $H^1(\R^N)$, and the conclusion of 
Lemma \ref{lem.estim-s} still holds true: there exists a unique solution 
of (\ref{eq5.1}) in $C\left([0,T];H^1(\R^N)\right)$ which is uniformly
bounded in $L^2\left((0,T)\times\R^N\right)$, independently of $\eps$.
Theorem \ref{thm.main} can be generalized as follows.

\begin{theorem}
\label{thm.em}
Assume (\ref{a1}) and (\ref{a2}) and that the initial data 
$u^0_\eps\in H^1(\R^N)$ is
$$
u^0_\eps(x) = \psi_n\left(\frac{x}{\eps},\theta^n\right)
e^{2i\pi\frac{\theta^n\cdot x}{\eps}} v^0(x) ,
$$
with $v^0\in H^1(\R^N)$.
The solution of (\ref{eq5.1}) can be written as 
$$
\begin{array}{l}
\dsp u_\eps(t,x) = e^{i\frac{\lambda_n(\theta^n)t}{\eps^{2}}}
e^{2i\pi\frac{\theta^n\cdot x}{\eps}} \psi_n\left(\frac{x}{\eps},\theta^n\right)
v_n(t,x) \\[0.2cm]
\dsp \phantom{u_\eps(t,x)} +
e^{i\frac{\lambda_m(\theta^m)t}{\eps^{2}}} 
e^{2i\pi\frac{\theta^m\cdot x}{\eps}} \psi_m\left(\frac{x}{\eps},\theta^m\right)
v_m(t,x) + r_\eps(t,x) ,
\end{array}
$$
with
$$
\lim_{\eps\to0} \int_0^T \int_{\R^N} \left| r_\eps(t,x) \right|^2 dx = 0 ,
$$
and $(v_n,v_m)\in C\left([0,T];L^2(\R^N)\right)^2$ is the unique solution
of the homogenized Schr\"{o}dinger system
\begin{equation}
\label{eq5.7}
\left\{ \begin{array}{ll}
\dsp i \frac{\partial v_n}{\partial t} - \div \left( A^*_n \nabla v_n \right) 
+ d^*_{nm}(t,x) \, v_m = 0 
& \mbox{ in } \R^N\times (0,T) \\[0.2cm]
\dsp i \frac{\partial v_m}{\partial t} - \div \left( A^*_m \nabla v_m \right) 
+ d^*_{mn}(t,x) \, v_n = 0
& \mbox{ in } \R^N\times (0,T) \\[0.2cm]
v_n(t=0,x) = v^0(x) & \mbox{ in } \R^N \\[0.2cm]
v_m(t=0,x) = 0 & \mbox{ in } \R^N ,
\end{array} \right.
\end{equation}
with $A^*_p= \frac{1}{8\pi^2} \nabla_\theta\nabla_\theta \lambda_p(\theta^p)$,
for $p=n,m$, and 
\begin{equation}
\label{eq5.7b}
\begin{array}{ll}
\dsp d^*_{nm}(t,x) = \overline d^*_{mn}(t,x) & = \dsp\frac i2
\int_{\T^N} \psi_m(y,\theta^m) a(t,x,y)\cdot
\left(\nabla-2i\pi\theta^n\right) \overline\psi_n(y,\theta^n) \, dy \\[0.3cm]
& \dsp - \frac i2 \int_{\T^N} \overline\psi_n(y,\theta^n) a(t,x,y)\cdot
\left(\nabla+2i\pi\theta^m\right) \psi_m(y,\theta^m) \, dy .
\end{array}
\end{equation}
\end{theorem}

\begin{remark}
In general $d^*_{nm}$ does not vanish, even if $a(t,x,y)$ is a
constant vector and $\theta^n=\theta^m$, so that there is indeed a
coupling between the two modes $n$ and $m$.
\end{remark}

\noindent
{\bf Proof of Theorem \ref{thm.em}.}
The proof is very similar to that of Theorem \ref{thm.main}. The first step
is identical, and in the second step we choose the same test function
$\Psi_\eps$, defined by (\ref{eq4.39}).
The higher order term in the variational formulation is
\begin{equation}
\label{eq5.8}
\begin{array}{ll}
\dsp \int_{\R^N} \left( \nabla +i\eps \AA_\eps \right) u_\eps \cdot
\left( \nabla -i\eps \AA_\eps \right)\overline \Psi_\eps dx
& \dsp = \int_{\R^N} \nabla u_\eps \cdot \nabla \overline\Psi_\eps \\[0.4cm]
& \dsp +i\eps \int_{\R^N} \left( u_\eps \AA_\eps \cdot\nabla\overline\Psi_\eps
- \overline\Psi_\eps \AA_\eps \cdot\nabla u_\eps \right) dx \\[0.4cm]
& \dsp + \eps^2 \int_{\R^N} |\AA_\eps|^2 u_\eps \overline\Psi_\eps \, dx .
\end{array}
\end{equation}
The first term in the right hand side of (\ref{eq5.8}) is exactly the
previous term (\ref{eq4.6}). The last one goes to zero, while the second 
one is the only new term which yields a non-zero limit. Indeed, integrating
by parts in this term gives
\begin{equation}
\label{eq5.9}
\begin{array}{l}
\dsp i\eps \int_{\R^N} \left( u_\eps \AA_\eps \cdot\nabla\overline\Psi_\eps
- \overline\Psi_\eps \AA_\eps \cdot\nabla u_\eps \right) dx
 = i\eps \int_{\R^N} u_\eps \left( 2 \AA_\eps \cdot\nabla\overline\Psi_\eps
+ \overline\Psi_\eps \div \AA_\eps \right) dx \\[0.4cm]
\dsp = i/2 \int_{\R^N} (v_\eps^{2n-m} +v_\eps^m) \overline\phi
\left( 2 a^\eps \cdot(\nabla_y-2i\pi\theta^n)\overline\psi_n^\eps
+ \overline\psi_n^\eps \div_y a^\eps \right) dx \\[0.4cm]
\dsp + i/2 \int_{\R^N} 2i\pi \left(  v_\eps^{2n-m}
(\theta^m-\theta^n)\cdot a^\eps + v_\eps^{m}
(\theta^n-\theta^m)\cdot a^\eps \right) \overline\phi
\overline\psi_n^\eps dx + {\cal O}(\eps) .
\end{array}
\end{equation}
Recalling that the two-scale limit of $v_\eps^{2n-m}$ is 0,
the limit of (\ref{eq5.9}) is
$$
i/2 \int_{\R^N} \int_{\T^N} v_m \overline\phi \psi_m
\left( 2a \cdot(\nabla_y-2i\pi\theta^n)\overline\psi_n
+ \overline\psi_n \div_y a + 2i\pi \overline\psi_n
(\theta^n-\theta^m)\cdot a \right) dx \, dy
$$
which yields formula (\ref{eq5.7b}) for the coupling coefficient
$d^*_{nm}$. The rest of the proof is identical to that of Theorem \ref{thm.main}.
$\Box$

\section{The resonant case}
\label{sec.reso}

In this section we come back to the original Schr\"{o}dinger equation 
(\ref{eq1.1}) but we change assumption (\ref{a2}) by assuming
that there is a single resonance between the initial data and the
target state, namely
\begin{equation}
\label{a2b}
\left\{ \begin{array}{ll}
(iii) & \mbox{ there exists } l\geq1 \mbox{ such that }
\lambda_l(2\theta^n-\theta^m)= 2\lambda_n(\theta^n)- \lambda_m(\theta^m) ,\\
(iv) & \mbox{ for any } p\geq1 , \quad
\lambda_p(3\theta^n-2\theta^m)\neq 3\lambda_n(\theta^n) - 2\lambda_m(\theta^m) .
\end{array} \right.
\end{equation}
We keep assumption (\ref{a1}) that we extend to the new
eigenvalue $\lambda_l$ for the Bloch parameter $\theta^l=2\theta^n-\theta^m$,
i.e.
\begin{equation}
\label{a1b}
\mbox{for } p=n,m,l 
\left\{ \begin{array}{ll}
(i) & \lambda_p(\theta^p) \mbox{ is a simple eigenvalue,} \\
(ii) & \theta^p \mbox{ is a critical point of } \lambda_p(\theta)
\mbox{ i.e., } \nabla_\theta\lambda_p(\theta^p) = 0 .
\end{array} \right.
\end{equation}
With these new assumptions we generalize Theorem \ref{thm.main}
by obtaining a limit system coupling three possible states
instead of just two.

\begin{theorem}
\label{thm.res}
Assume (\ref{a1b}) and (\ref{a2b}) and that the initial data $u^0_\eps\in H^1(\R^N)$
is
$$
u^0_\eps(x) = \psi_n\left(\frac{x}{\eps},\theta^n\right)
e^{2i\pi\frac{\theta^n\cdot x}{\eps}} v^0(x) ,
$$
with $v^0\in H^1(\R^N)$.
The solution of (\ref{eq1.1}) can be written as
$$
\begin{array}{l}
\dsp u_\eps(t,x) = e^{i\frac{\lambda_n(\theta^n)t}{\eps^{2}}}
e^{2i\pi\frac{\theta^n\cdot x}{\eps}} \psi_n\left(\frac{x}{\eps},\theta^n\right)
v_n(t,x) \\[0.2cm]
\dsp \phantom{u_\eps(t,x)} +
e^{i\frac{\lambda_m(\theta^m)t}{\eps^{2}}}
e^{2i\pi\frac{\theta^m\cdot x}{\eps}} \psi_m\left(\frac{x}{\eps},\theta^m\right)
v_m(t,x) \\[0.2cm]
\dsp \phantom{u_\eps(t,x)} +
e^{i\frac{\lambda_l(\theta^l)t}{\eps^{2}}}
e^{2i\pi\frac{\theta^l\cdot x}{\eps}} \psi_l\left(\frac{x}{\eps},\theta^l\right)
v_l(t,x) + r_\eps(t,x) ,
\end{array}
$$
with
$$
\lim_{\eps\to0} \int_0^T \int_{\R^N} \left| r_\eps(t,x) \right|^2 dx = 0 ,
$$
and $(v_n,v_m,v_l)\in C\left([0,T];L^2(\R^N)\right)^3$ is the unique solution
of the homogenized Schr\"{o}dinger system
$$
\left\{ \begin{array}{ll}
\dsp i \frac{\partial v_n}{\partial t} - \div \left( A^*_n \nabla v_n \right)
+ d^*_{nm}(t,x) \, v_m + d^*_{nl}(t,x) \, v_l = 0
& \mbox{ in } \R^N\times (0,T) \\[0.3cm]
\dsp i \frac{\partial v_m}{\partial t} - \div \left( A^*_m \nabla v_m \right)
+ d^*_{mn}(t,x) \, v_n = 0
& \mbox{ in } \R^N\times (0,T) \\[0.3cm]
\dsp i \frac{\partial v_l}{\partial t} - \div \left( A^*_l \nabla v_l \right)
+ d^*_{ln}(t,x) \, v_n = 0
& \mbox{ in } \R^N\times (0,T)  \\[0.2cm]
v_n(t=0,x) = v^0(x) & \mbox{ in } \R^N \\[0.2cm]
v_m(t=0,x) = 0 & \mbox{ in } \R^N  \\[0.2cm]
v_l(t=0,x) = 0 & \mbox{ in } \R^N ,
\end{array} \right.
$$
with $A^*_p= \frac{1}{8\pi^2} \nabla_\theta\nabla_\theta \lambda_p(\theta^p)$,
for $p=n,m,l$, and
\begin{equation}
\label{eq7.7b}
d^*_{np}(t,x) = \overline d^*_{pn}(t,x) = \frac12
\int_{\T^N} d(t,x,y)\overline\psi_n(y,\theta^n)\psi_p(y,\theta^p) \, dy
\end{equation}
for $p=m,l$.
\end{theorem}

\begin{remark}
More generally, there could be multiple resonances between the initial
and target state. Let $k_0\geq1$ be the order of the resonance. Under
a suitable generalization of assumption (\ref{a2b}), all modes of
momentum $(k+1)\theta^n-k\theta^m$ and energy
$(k+1)\lambda_n(\theta^n)-k\lambda_m(\theta^m)$ are coupled for
$-1\leq k\leq k_0$. Theorem \ref{thm.res} can be generalized to
obtain an homogenized system for $(v_m,v_n,v_{2n-m},
..., v_{(k_0+1)n-k_0m})$ in which the coupling matrix $d^*$ is
hermitian of size $k_0+2$ with the following sparse structure
$$
d^* = \left( \begin{array}{ccccccc}
0 & \times & 0 & & & & \\
\times & 0 & \times & 0 &  & & \\
0 & \times & 0 & \times & 0 & &  \\
 & \ddots & \ddots & \ddots & \ddots & \ddots & \\
&&0&\times & 0 & \times & 0 \\
&&&0 & \times & 0 & \times \\
&&&&0& \times & 0
\end{array} \right)
$$
\end{remark}

\noindent
{\bf Proof of Theorem \ref{thm.res}.}
The only modification with respect to the proof of Theorem \ref{thm.main}
is the fact that the sequence $v_\eps^{2n-m}$, defined by (\ref{eq4.23}),
now admits a non-zero two-scale limit $\psi_l(y,\theta^l) v_l(t,x)$
because the spectral cell problem (\ref{eq4.9b}) has a non-trivial
solution $\psi_l$, as a consequence of part $(iii)$ of assumption
(\ref{a2b}). No other states appear because of part $(iv)$ in (\ref{a2b}).
The rest of the proof is similar to that of Theorem \ref{thm.main}
and we safely leave it to the reader.
$\Box$ 

\bigskip

\noindent
{\bf Acknowledgments.}  This work was partly done when M. Vanninathan
was visiting the Centre de Math\'ematiques Appliqu\'ees
at Ecole Polytechnique. The support of the MULTIMAT european network
MRTN-CT-2004-505226 is kindly acknowledged by G. Allaire.
The authors thank G. Milton for bringing
this problem to their attention.

\end{document}